# Misleading reference to unpublished wound ballistics data regarding distant injuries

Michael Courtney, PhD and Amy Courtney, PhD
Ballistics Testing Group, P.O. Box 24 West Point, NY 10996
Michael_Courtney@alum.mit.edu, Amy_Courtney@post.harvard.edu

**Abstract:** An article (*J Trauma* 29:10-18, 1989) cites unpublished wound ballistics data to support the authors' view that distant injuries are a myth in wound ballistics. The actual data, published in 1990, actually contains a number of detailed examples of distant injuries. (Bellamy RF, Zajtchuk R. The physics and biophysics of wound ballistics. In: Zajtchuk R, ed. *Textbook of Military Medicine, Part I: Warfare, Weaponry, and the Casualty, Vol. 5, Conventional Warfare: Ballistic, Blast, and Burn Injuries.* Washington, DC: Office of the Surgeon General, Department of the Army, United States of America; 1990: 107-162)

Is there a statute of limitations on apparent misrepresentations of unpublished data and personal communications? In a letter to the editor disputing the findings of Suneson et al. (*J Trauma* 29:10-18, 1989) related to distant injuries ascribed to ballistic pressure waves, ML Fackler and CE Peters claim (*J Trauma* 29:1455; 1989):

*A review of 1400 rifle wounds from Vietnam (Wound Data and Munitions Effectiveness Team) should lay to rest the myth of "distant" injuries. In that study, there were no cases of bones being broken, or major vessels torn, that were not hit by the penetrating bullet. In only two cases, an organ that was not hit (but was within a few cm of the projectile path), suffered some disruption (personal communication, Bellamy, R.F., 1989).*

In contrast to these claims, Bellamy's published analysis of the data the following year[1] describes a number of cases of distant wounding including broken bones (pp. 153-154), five instances of abdominal wounding in cases where the bullet did not penetrate the abdominal cavity (pp. 149-152), a case of lung contusion resulting from a bullet hit to the shoulder (pp. 146-149), and a case of distant effects on the central nervous system (p. 155).

The errant letter to the editor and related claim that "the review of 1400 rifle wounds" disproves distant injuries has been occasionally cited[2,3,4] even though the published analysis[1] actually supports findings of distant injury. The Suneson et al. studies related to remote CNS damage[5-9] have been favorably cited by a number of authors and independently confirmed.[10] Distant effects of penetrating projectiles have also found support in a number of other studies[11-16].

References:

1. Bellamy RF, Zajtchuk R. The physics and biophysics of wound ballistics. In: Zajtchuk R, ed. *Textbook of Military Medicine, Part I: Warfare, Weaponry, and the Casualty, Vol. 5, Conventional Warfare: Ballistic, Blast, and Burn Injuries.* Washington, DC: Office of the Surgeon General, Department of the Army, United States of America; 1990: 107-162.

2 Barlett CS, Bissell BT. Common misconceptions and controversies regarding ballistics and gunshot wounds. *Tech. in Orthop.* 2006; 21: 190-199.

3 Bartlett CS. Clinical update: gunshot wound ballistics. *Clin. Orthop.* 2003; 408: 28-57.

4. Hollerman JJ. Wound ballistics is a model of the pathophysiology of all blunt and penetrating trauma. Emergency Radiology. 1998; 5: 279-288.

5. Suneson A, Hansson HA, Kjellström BT, Lycke E, and Seemn T. Pressure waves by high energy missile impair respiration of cultured dorsal root ganglion cells. *J Trauma.* 1990; 30:484-488.

6. Suneson A, Hansson HA, Seeman T: Peripheral high-energy missile hits cause pressure changes and damage to the nervous system: experimental studies on pigs. *J Trauma.* 1987; 27:782-789.

7. Suneson A, Hansson HA, Seeman T: Central and peripheral nervous damage following high-energy missile wounds in the thigh. *J Trauma.* 1988; 28:S197-S203.

8. Suneson A, Hansson HA, Seeman T: Pressure wave injuries to the nervous system caused by high energy missile extremity impact: part I. local and distant effects on the peripheral nervous system. A light and electron microscopic study on pigs. *J Trauma.* 1990; 30:281-294.

9. Suneson A, Hansson HA, Seeman T: Pressure wave injuries to the nervous system caused by high energy missile extremity impact: part II. distant effects on the central nervous system. A light and electron microscopic study on pigs. *J Trauma.* 1990; 30:295-306.

10. Wang Q, Wang Z, Zhu P, Jiang J. Alterations of the myelin basic protein and ultrastructure in the limbic system and the early stage of trauma-related stress disorder in dogs. *J Trauma.* 2004;56:604-610.

11. Treib J, Haass A, Grauer MT. High-velocity bullet causing indirect trauma to the brain and symptomatic epilepsy. *Mil. Med.* 1996;161:61-64.

12. Sturtevant B. Shock wave effects in biomechanics. *Sadhana.* 1998; 23: 579-596.

13. Tien HC, van der Hurk TWG, Dunlop P, Kropelin B, Nahouraii R, Battad AB, van Egmond T. Small bowel injury from a tangential gunshot wound without peritoneal penetration: a case report. *J Trauma*. 2007;62:762-762.

14. Gryth D, Rocksen D, Persson JKE, Arborelius WP, Drobin D, Bursell J, Olsson LG, Kjellstrom TB. Severe lung contusion and death after high-velocity behind-armor blunt trauma: relation to protection level. *Mil. Med.* 2007;172:1110-1116.

15. Merkle AC, Ward ES, O'Conner JV, Roberts J. Assessing behind armor blunt trauma (BABT) under NIJ standard-0101.04 conditions using human torso models. *J Trauma*. 2008; 64:1555-1561.

16. Courtney A, Courtney M. Links between traumatic brain injury and ballistic pressure waves originating in the thoracic cavity and extremities. *Brain Injury.* 2007; 21: 657-662.